\documentclass[12 pt]{article}
\usepackage{graphicx}
\usepackage{amssymb}
\begin{document}
\title{Peculiarities of anharmonic effects in the lattice thermodynamics
of fcc metals}
\date{}
\author{M. I. Katsnelson, A. F. Maksyutov$^\dag$, A. V. Trefilov$^\dag$}
\maketitle
\begin{center}
Institute of Metal Physics, 620219 Ekaterinburg, Russia
\\
$^\dag$ Russian Research Centre "Kurchatov Institute", 123182 Moscow,
Russia
\\
Corresponding author: M. I. Katsnelson, e-mail: Mikhail.Katsnelson@usu.ru
\end{center}

\begin{abstract}
Explicit expressions for anharmonic contributions to the thermodynamic
properties with allowance for higher-order phonon-phonon interactions for
closed-packed crystals are given, and the calculations for some fcc
metals near the melting (Ir, Rh) and martensite phase transition (Ca, Sr)
points are carried out. A detailed comparison of anharmonic and electron
contributions to the heat capacity of these metals is carried out. The
computational results for high-temperature heat capacity agree well with
the available experimental data.
\end{abstract}



Anharmonic effects (AE) in the lattice dynamics and thermodynamics have
been extensively discussed in the literature [1-5]. However the
quantitative information on AE, both experimental and theoretical, is
rather poor so far. The experimental difficulties are due to the fact that at
high temperatures ($T\sim T_m$, where $T_m$ is the melting temperature) when the
AE in the thermodynamic properties become noticeable, it is usually
difficult to identify their contribution to the heat capacity and thermal
expansion from the contributions of vacancies and other thermally excited
lattice defects [2,5-8]. The theoretical calculations, on the other hand,
are based, as a rule, on rough models and, therefore, do not give any
definite information on the AE in thermodynamic of real crystals. At the
same time this information is needed for understanding the high
temperature properties of solids. For example, it is important to clear
up what is the role of self-anharmonic effects associated with
phonon-phonon interactions, and what is the role of quasiharmonic ones
which are only due to the change in the volume because of thermal
expansion. Simple estimations \cite{pai} show that the AE contribution to
heat capacity at high temperatures of the order of the electron
contribution to heat capacity and has the same (linear) temperature
dependence; the real ratio of these two contributions essentially depend
on the features of the electron structure near the Fermi level $E_F$.
However this problem has not been studied yet. All these problems are of
special importance for refractory metals ($T_m>2000$~K) and for the metals
with high-temperature martensite transitions which are usually
accompanied by essential increase in the AE in the lattice dynamics
\cite{KTKK}. In the present work the AE features (in the comparison with
the electron contributions) in the thermodynamical properties of
refractory fcc metals Ir and Rh and in the bcc and fcc phases of Ca and
Sr near the points of structural transformation are studied. The
calculations were made in terms of microscopic models used in
\cite{KTKK,JL} which describe a wide range of the lattice properties of
these metals.

The initial Hamiltonian of phonon subsystem is
\begin{equation}
H=\sum_\lambda \left (\frac{P_{\lambda}^2}{2M} + \frac{M\omega_\lambda^2
Q_\lambda^2}{2} \right ) +\sum_{n=3}^\infty H^{(n)}
\label {eq1}
\end{equation}
\begin{equation}
H^{(n)}=\sum_{\lambda_1\dots\lambda_n}
\frac{
   \Phi^{(n)} (\lambda_1,\dots,\lambda_n)
}
{n!}
Q_{\lambda_1}\dots Q_{\lambda_n}
\label{eq2}
\end{equation}
where $M$ is the mass of atoms; $\omega_\lambda$ is the phonon frequency,
$\lambda\equiv{\bf q}\xi$ where ${\bf q}$~-- is the wave vector; $\xi$ is
the branch number, $Q_\lambda$, $P_\lambda$ are the normal coordinates in
the harmonic approximation and the corresponding pulses, $\Phi^{(n)}$ are
amplitudes of intermode interactions. In the leading order of the
parameter $\eta\equiv T/E_{at}$ where $E_{at}$ is the energy of the order
of binding energy, the basic contribution of AE to free energy $F$ is
determined by the second order by $H^{(3)}$ and the first order by
$H^{(4)}$ \cite{Cow,VKT}
\begin{equation}
F_{an}=F^{(3)}+F^{(4)}
\label{eq3}
\end{equation}
where at $T\gg\Theta_D$ ($\Theta_D$ is the Debye temperature),
\begin{equation}
F^{(3)}=- \frac{T^2}{12 M^3} \sum_{\lambda\mu\nu} \frac{|\Phi^{(3)}(\lambda,
\mu,\nu)|^2}{\omega_\lambda^2\omega_\mu^2\omega_\nu^2},
\label{eq4}
\end{equation}
\begin{equation}
F^{(4)}=\frac{T^2}{4 M^2} \sum_{\lambda\mu} \frac{\Phi^{(4)}(\lambda,
\lambda,\mu,\mu)}{\omega_\lambda^2\omega_\mu^2};
\label{eq5}
\end{equation}
as usually, summing over phonon quasimomenta is fulfilled taking into
account the conservation law for $\lambda\equiv(\xi,{\bf k})$,
$\mu\equiv(\eta,{\bf q})$, $\nu\equiv(\zeta,{\bf k+q})$. The explicit
microscopic expressions for $\Phi^{(3)}$ and $\Phi^{(4)}$ are given in
\cite{VKT}.

The heat capacity of the crystal at $T\gg\Theta_D$, without the standard
harmonic contribution $C_V^{harm}=3R$ (where $R$ is the gaseous
constant), has the form:
\begin{equation}
\Delta C=C_P-C_V^{harm}=(C_P-C_V)+C_V^{an}+C_V^{e}
+C_V^{d}
\label{eq6}
\end{equation}
where $C_V^{an}$ are the anharmonic contributions to the heat capacity
\begin{equation}
C_V^e(T)=\frac{R}{T}\int\limits_{-\infty}^\infty {\rm d}E
\left [-\frac{\partial f(E)}{\partial E} \right ]
(E-\mu)^2 N(E)
\label{eq7}
\end{equation}
is the electron heat capacity ($N(E)$ is the electron density of states,
is the chemical potential, $f(E)$ is the Fermi distribution function), $C_V^d$ is
the contribution of defects; the calculation formula for $C_P-C_V$ in the
quasiharmonic approximation is presented in \cite{VKT}. The density of
states $N(E)$ was calculated by the FP-LMTO method in the local
approximation for the density functional (LDA) \cite{savras}.
To estimate the contribution of defects to the heat capacity a common
approximation of independent monovacancies was used with
\begin{equation}
C_V^d(T)=R\left(\frac{E_v}{T}\right)^2
\exp\left(S_v-\frac{E_v}{T}\right)
\label{eq71}
\end{equation}
where $E_v$, $S_v$ are the energy and entropy of monovacancy formation,
respectively.

The calculations of AE in the Ir lattice dynamics, made in \cite{JL}, show
that near the melting temperature $T=T_m$ the temperature dependences of
the phonon frequency and phonon damping drastically change over the
Brillouin zone reaching the values of the order of 15-20\%. This value is
high enough so that the question on the role of higher orders in the
anharmonic perturbation theory could be set. The theory of
self-consistent phonons, often used for this purpose (see, e.g.,
\cite{betger}) is not adequate in this case as it only accounts for a
part of $H^{(2n)}$ contributions to (\ref{eq1}) and neglects the
$H^{(2n+1)}$ contributions. At the same time, the calculations \cite{JL}
show that in Ir the contributions of three-phonon processes to the
temperature shifts of the phonon frequencies always dominate over the contribution
of four-phonon processes. In the recent work \cite{shukla} another
approximation is proposed which accounts both the three-phonon and
four-phonon processes, and, as stated, reasonably describes the
thermodynamics of crystals with the Lennard-Jones type interaction. The
applicability of this approximation in the case considered seems somewhat
doubtful as it is based on the replacement of anharmonic frequency shift
by a certain value average over the Brillouin zone while the calculations
in \cite{JL} show that the {\bf q}-dependence of AE is essential.
Moreover, it is known that the approximations adequate for the
description of thermodynamics of systems with ``hard'' potentials of
Lennard-Jones type are not usually applicable for metals as they have
relatively ``soft'' potentials \cite{VKT}.

To calculate the effects of higher orders by the Hamiltonian $H^{(3)}$
the approximation similar to that of Gell-Mann and Brueckner (``bubble
approximation'') in the theory of electron liquid \cite{GB} was used. It can be
shown that  the accurate expression for the contribution of three-phonon
processes to the free energy at $T\gg\Theta_D$ has the form:
\begin{equation}
F^{(3)}=-\frac{T}{3} \int\limits_0^1 \frac{d\alpha}{\alpha}
\sum_\lambda \left [G_\lambda(\omega=0)\Sigma_\lambda(\omega=0)
\right ]\biggr|_{H^{(3)}\to \alpha H^{(3)}}
\label{eq8}
\end{equation}
where $G_\lambda$, $\Sigma_\lambda$ are the Green function and
self-energy part for phonon (see \cite{Cow}) connected by the Dyson equation

\begin{equation}
G_\lambda^{-1}(\omega)=\frac{\omega^2-\omega_\lambda^2}{2\omega_\lambda}-\Sigma_\lambda(\omega).
\label{eq9}
\end{equation}
The calculations \cite{JL} show that the AE in fcc metals have pronounsed
peaks as functions of ${\bf q}$ at points X and L. The largest
contribution to the free energy comes from the ``bubble'' diagrams; this
situation is similar to the electron gas of high density with the only
difference that in the latter case the ``dangerous'' region is ${\bf q}\approx0$
\cite{GB}. Taking this into account, in (\ref{eq8})-(\ref{eq9}) we may
restrict ourselves to the expression of the lowest (second) order of the
perturbation theory for $\Sigma_\lambda$. Then we have
\begin{equation}
F^{(3)}=\frac{T}{6}\sum_\lambda \ln \left [1-\frac{T}{2 M^3}
\sum_{\mu\nu}
\frac{|\Phi^{(3)}(\lambda\mu\nu)|^2}{\omega_\lambda^2\omega_\mu^2\omega_\nu^2}
\right ]
\label{eq10}
\end{equation}

Unlike the bcc metals the fcc metals have no soft modes in the phonon
spectrum, which makes the four-phonon processes relatively small
comparing with the three-phonon ones \cite{KTKK,JL}. Therefore for $F^{(4)}$ it is
sufficient to use the simplest corresponding to the Hartry-Fock
approximation (the simplest procedure of disengagement in averaging
Hamiltonian $H^{(4)}$). Then for $T\gg\Theta_D$ we have
\begin{equation}
F^{(4)}=T\sum_\lambda \left [\ln \left (1+\frac{\Delta_{4\lambda}}
{\omega_\lambda} \right ) -\frac{1}{2}\frac{\Delta_{4\lambda}}
{\omega_\lambda} \right ],
\label{eq11}
\end{equation}
where
\begin{equation}
\Delta_{4\lambda} =\frac{T}{4 \omega_\lambda} \sum_\mu \
\frac{\Phi^{(4)} (\lambda,\lambda,\mu,\mu)} {M^2 \omega_\mu^2}
\label{eq12}
\end{equation}
is the corresponding correction to the phonon frequency.

\begin{figure}[htp]
\begin{center}
\includegraphics[height=110mm,width=80mm,angle=270]{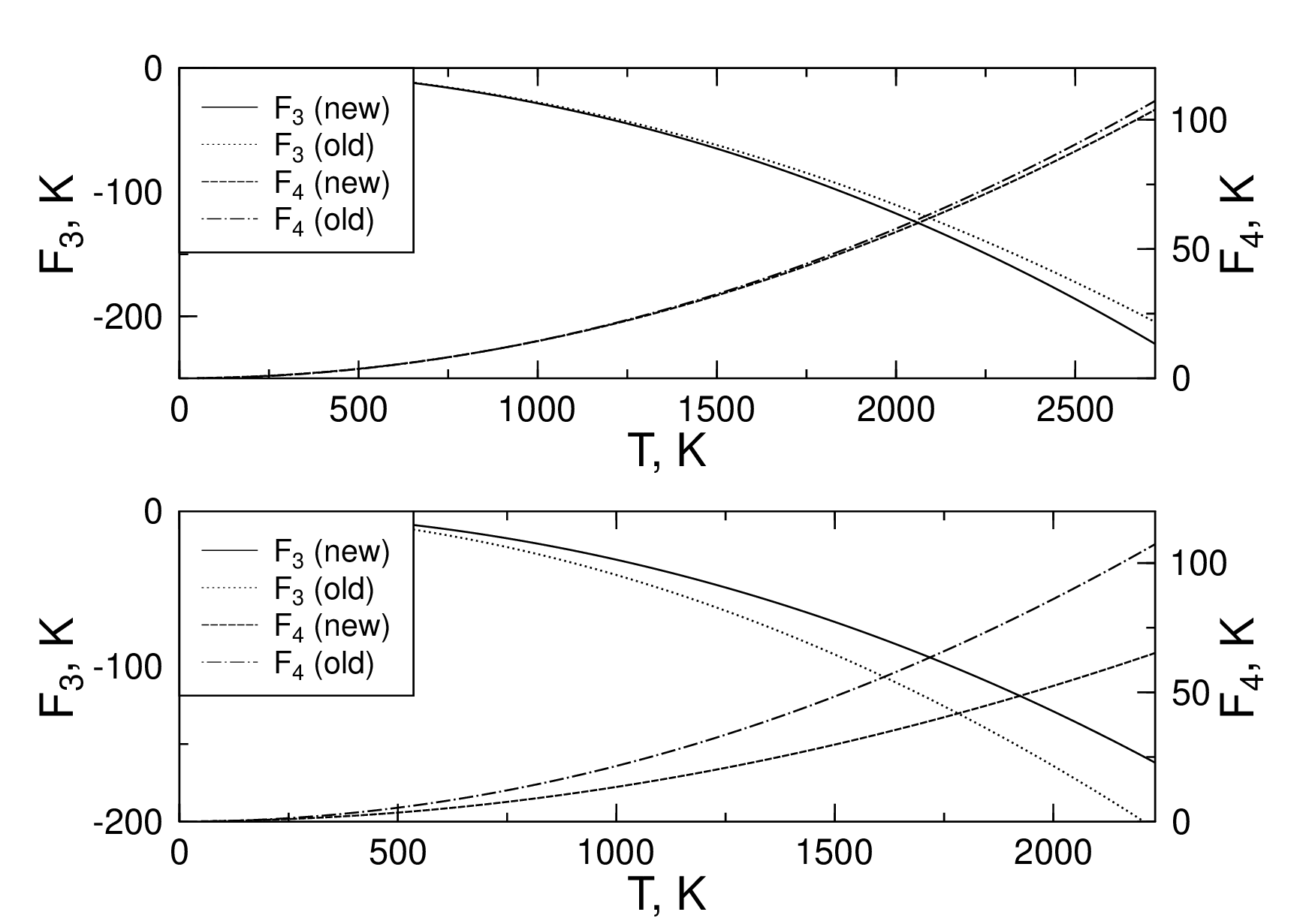}
\end{center}
\vskip -3mm
{\footnotesize
Fig. 1. Anharmonic contributions to the free energy of Ir and Rh (in K)
considered by the perturbation theory (\ref{eq4}), (\ref{eq5})
(``old'') and with taking into account the higher orders (\ref{eq10}),
(\ref{eq11}) (``new'').}
\vskip 2mm
\end{figure}
\begin{figure}[htp]
\begin{center}
\includegraphics[height=110mm,width=80mm,angle=270]{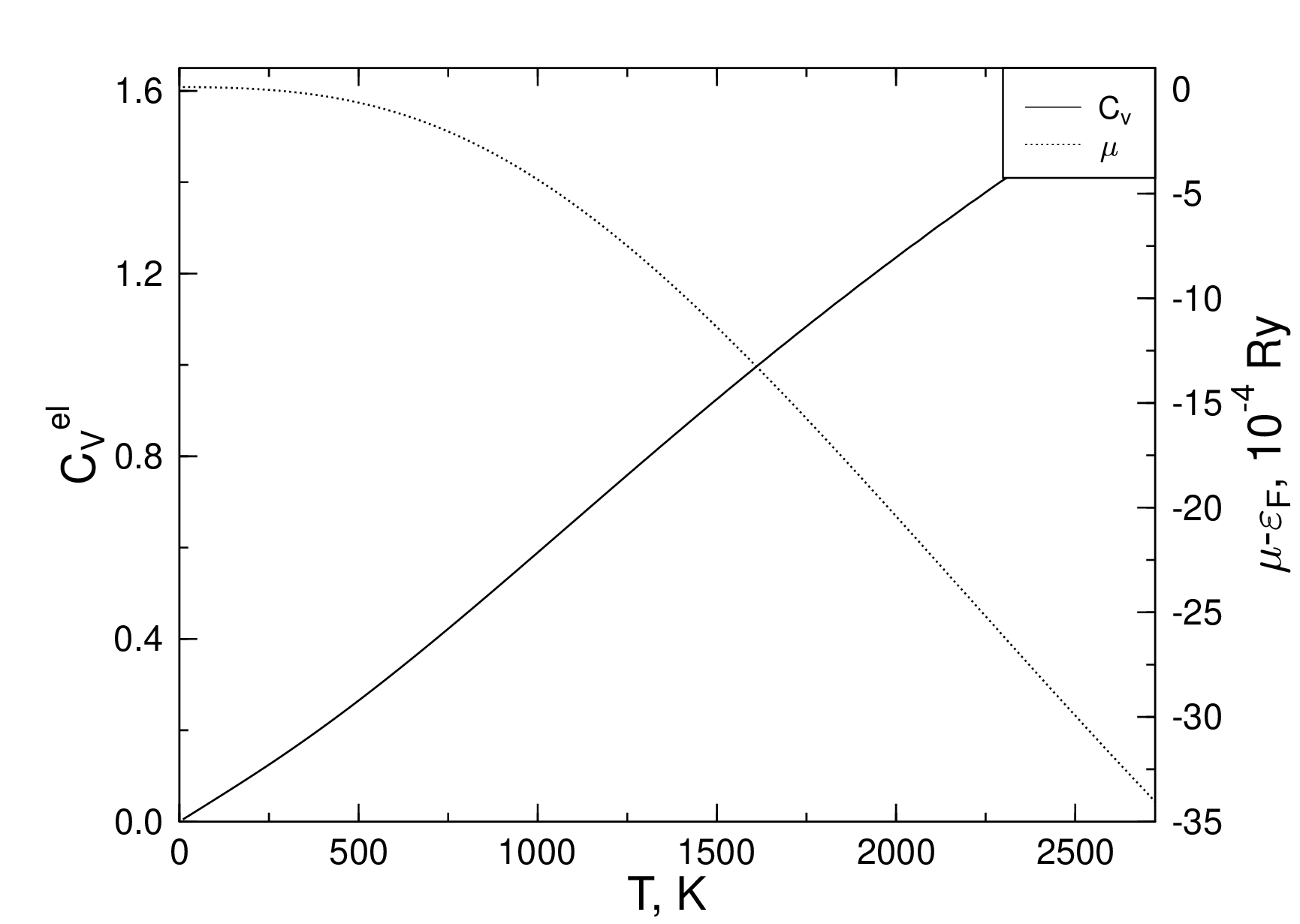}
\end{center}
\vskip -3mm
{\footnotesize
Fig. 2. Electron contribution to the heat capacity (in $R$ units)
and temperature dependence of chemical potential (to $10^{-4}$~Ry) in Ir;
the dependencies for Rh are similar.}
\vskip 2mm
\end{figure}
\begin{figure}[htp]
\begin{center}
\includegraphics[height=110mm,width=80mm,angle=270]{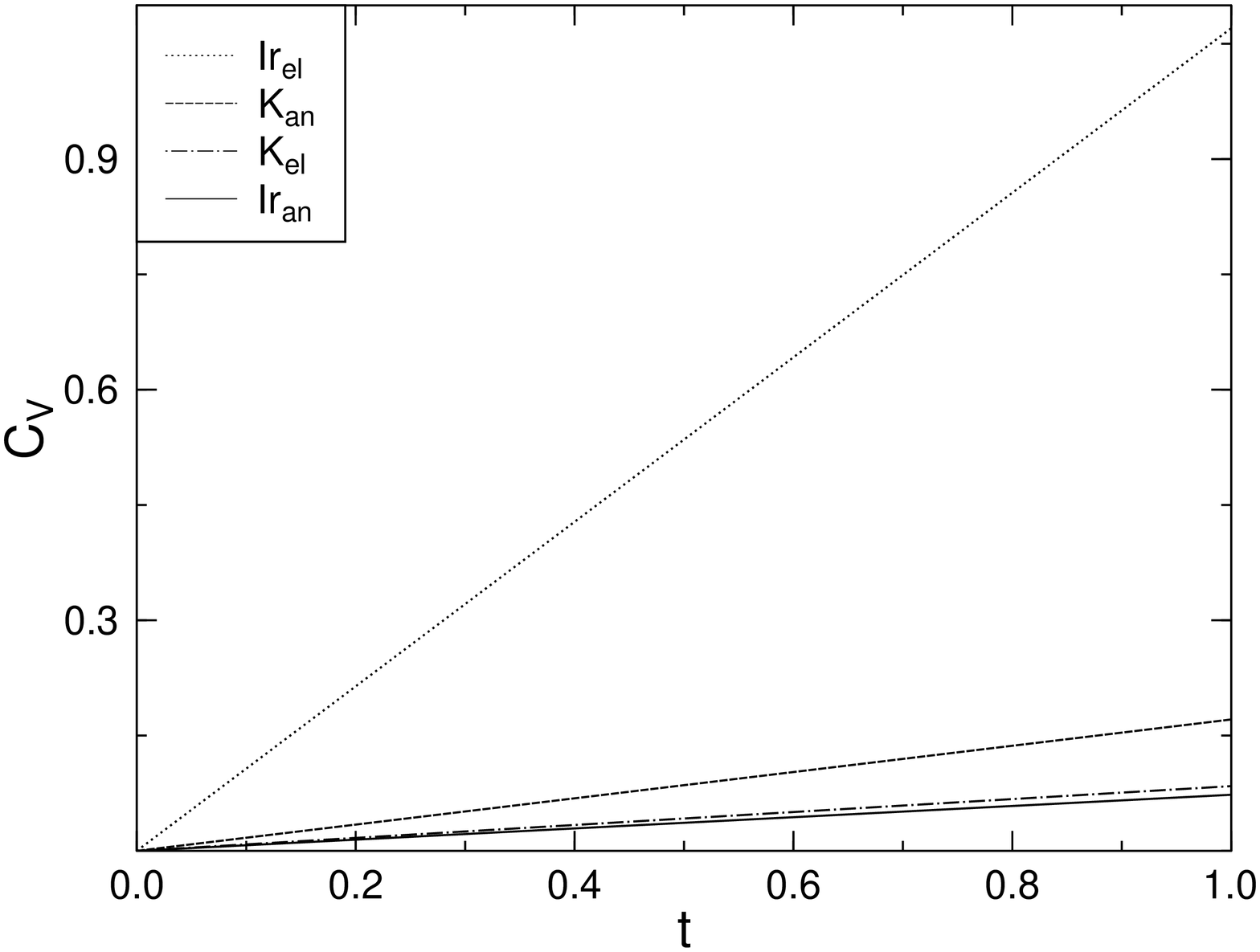}
\end{center}
\vskip -3mm
{\footnotesize
Fig. 3. Comparison of the electron and anharmonic contributions to
the heat capacity (in $R$ units) for the fcc phases of K and Ir; $t=T/T_m$.}
\vskip 2mm
\end{figure}
\begin{figure}[htp]
\begin{center}
\includegraphics[height=110mm,width=80mm,angle=270]{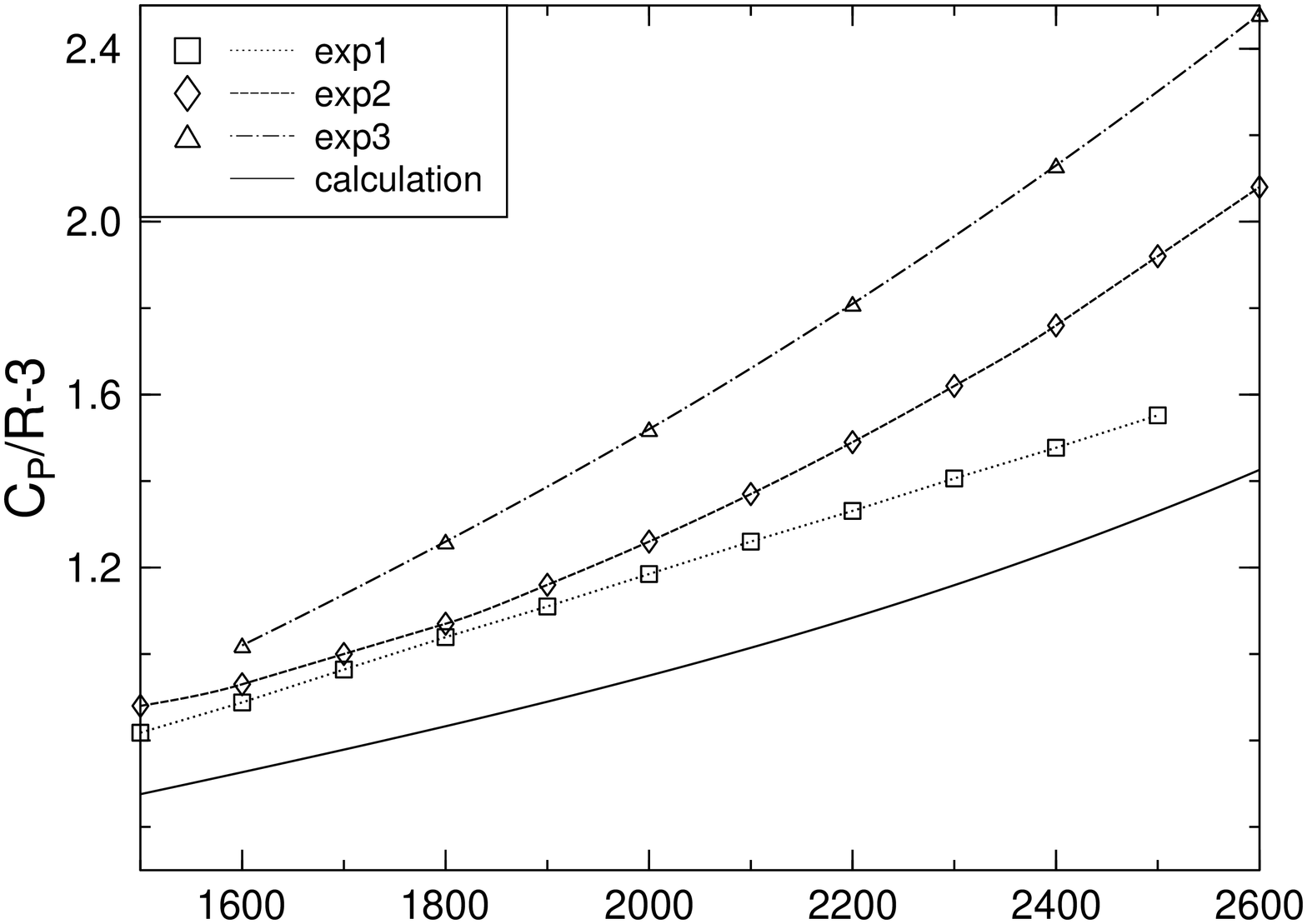}
\end{center}
\vskip -3mm
{\footnotesize
Fig. 4. Comparison of the calculated values of heat capacity $C_P$
(formula (\ref{eq6})) with the experiment for Ir; the data of  exp1, exp2,
exp3 are taken from \cite{truh, ramana, kats}, respectively.}
\vskip 2mm
\end{figure}
\begin{figure}[htp]
\begin{center}
\includegraphics[height=110mm,width=80mm,angle=270]{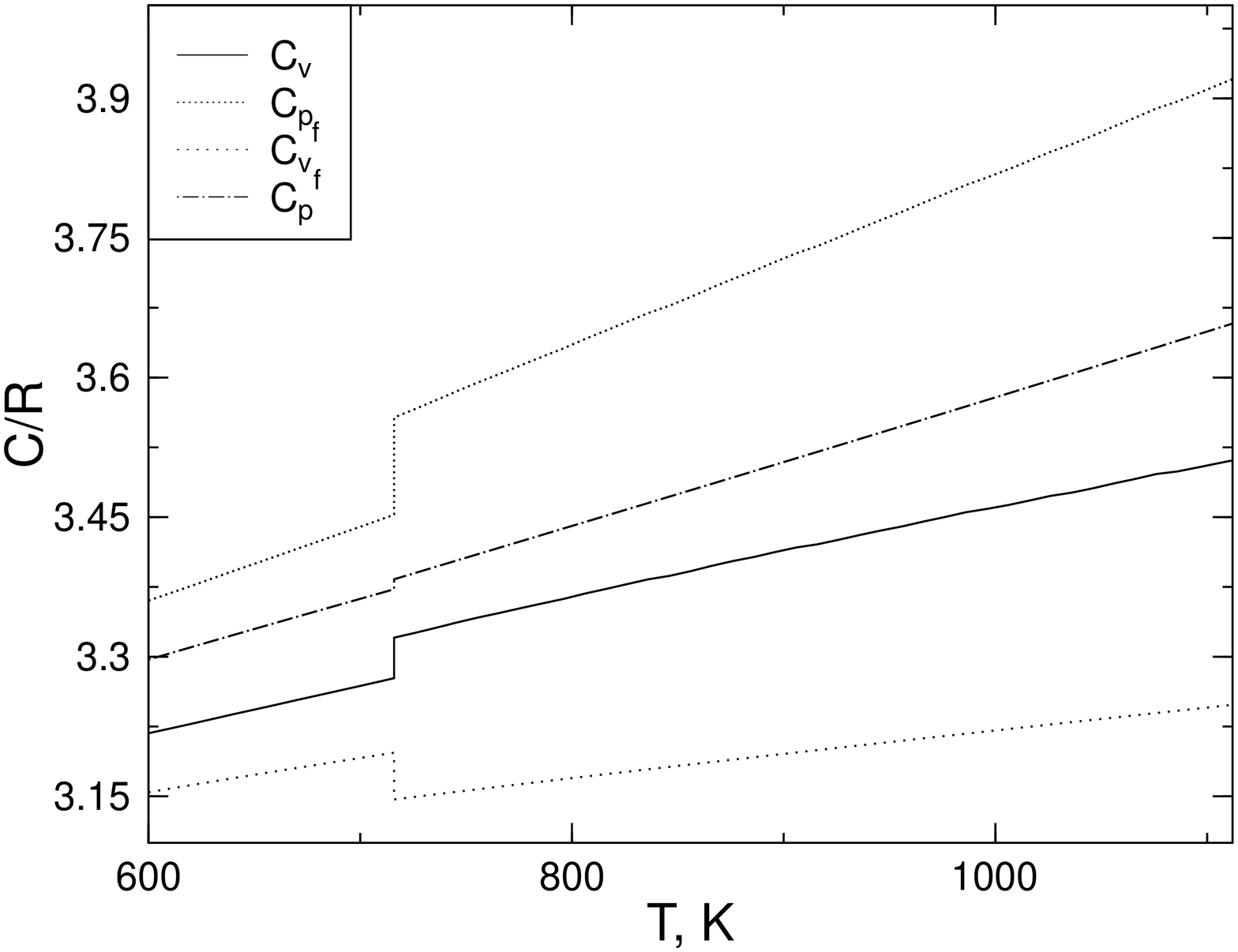}
\end{center}
\vskip -3mm
{\footnotesize
Fig. 5. Theoretical value of heat-capacity in Ca. The notation
introduced $C_v=C_v^{harm}+C_v^{an}+C_v^{el}$,
$C_p=C_v^{harm}+C_v^{an}+C_v^{el}+(C_p-C_v)$,
$C_v^f=C_v^{harm}+C_v^{an}+C_v^{free\ el}$,
$C_p^f=C_v^{harm}+C_v^{an}+C_v^{free\ el}+(C_p-C_v)$,
$C_v^{free\ el}$~-- the electron heat capacity in the approximation of
free electrons.}
\vskip 2mm
\end{figure}

The computational results are shown in Figs.~1-5. One can see that the AE
higher order contributions to the free energy of Ir and Rh are not large
(do not exceed 20\%). In Ir and Rh the anharmonic contribution to heat
capacity is small comparing with the electron contribution while in the
sp-metals (as the hypothetical fcc phase K shown in Fig.~3) these
contributions are comparable. Among the metals studied only for Ir the
detailed experimental information is available, the theoretical data
being in good agreement with these results (Fig.~4). In the estimation of
$C_V^d$ according to (\ref{eq71}) the the first principles calculation
result from  $E_v=1.88$~eV \cite{GKM} was used for $E_v$, and $S_v=2.5$ for
$S_v$ \cite{VKT}. It follows from Fig.~5 that in the difference of
heat-capacity for bcc and fcc phases of Ca (and similarly Sr) a strong
compensation for different contributions -- electron, anharmonic ones --
and the contribution of $C_P-C_V$ takes place. Thus, for the discussion
of thermodynamics of high temperature structure transformations in metals
the microscopic calculations of corresponding values including the
anharmonic effects are needed.

The authors are grateful to K.~Yu.~Khromov for his assistance in
performance of calculations and M.~N.~Khlopkin for helpful discussion of
experimental results.

The work is sponsored by the Russian Basic Research Foundation, grant
01-02-16108, and the Netherlands Research Organization, grant
NWO 047-008-16.

\end{document}